\documentclass[prb,twocolumn,showpacs,preprintnumbers,amsmath,amssymb,floatfix]{revtex4}
\usepackage{graphicx}% Include figure files
\usepackage{float}
\usepackage{trfsigns}

\newcommand \be{\begin{eqnarray}}
\newcommand \ee{\end{eqnarray}}

\newcommand \ba{\begin{align}}
\newcommand \eea{\end{align}}

\newcommand \V{\vec}
\newcommand {\ov}[1]{\overline{#1}}

\begin{document}
%\twocolumn[\hsize\textwidth\columnwidth\hsize
           \csname @twocolumnfalse\endcsname
\title{Conditions where RPA becomes exact in the high-density limit}
\author{Klaus Morawetz$^{1,2,3}$,
%Paul Ziesche $^3$,
 Vinod Ashokan$^4$, Renu Bala$^5$,
%Dheeraj Kumar Shukla$^5$,
Kare Narain Pathak$^4$
}
\affiliation{$^1$M\"unster University of Applied Sciences,
Stegerwaldstrasse 39, 48565 Steinfurt, Germany}
\affiliation{$^2$International Institute of Physics- UFRN,
Campus Universit\'ario Lagoa nova,
59078-970 Natal, Brazil
}
\affiliation{$^{3}$ Max-Planck-Institute for the Physics of Complex Systems, 01187 Dresden, Germany}
\affiliation{$^4$ Centre for Advanced Study in Physics, Panjab University, 160014 Chandigarh, India}
\affiliation{$^5$ Department of Physics, MCM DAV College for Women, 160036 Chandigarh, India}

\begin{abstract}
It is shown that in $d$-dimensional systems, the vertex corrections beyond the random phase approximation (RPA) or GW approximation scales with the power $d-\beta-\alpha$ of the Fermi momentum if the relation between Fermi energy and Fermi momentum is $\epsilon_{\rm f}\sim p_{\rm f}^\beta$ and the interacting potential possesses a momentum-power-law of $\sim p^{-\alpha}$. The condition $d-\beta-\alpha<0$ specifies systems where RPA is exact in the high-density limit. The one-dimensional structure factor is found to be the interaction-free one in the high-density limit for contact interaction. A cancellation of RPA and vertex corrections render this result valid up to second-order in contact interaction. For finite-range potentials of cylindrical wires a large-scale cancellation appears and found to be independent of the width parameter of the wire. The proposed high-density expansion agrees with the Quantum Monte Carlo simulations. 
\end{abstract}
\maketitle
%    \vskip2pc]

The correlation energy in electron gases has been a topic of long-time investigations. In order to avoid divergences in perturbation theory already Macke \cite{M50} summed an infinite series of diagrams (RPA). Later Gell-Man and Bruckner show that the RPA at zero temperature becomes exact in the high-density limit \cite{G57,Mar58} confirmed up to orders of the logarithm of density \cite{WP91}. The corresponding momentum distributions in RPA have been computed already in \cite{DV60,K61} and recently an improved parametrization has been presented by cummulant expansions \cite{PZ02}. It confirms that the high-density limit is indeed given by the RPA calculation. The analytic expressions of the electron gas have been found \cite{Zi07,Z10} and an approximation bridging the low and high-density expansion of the correlation energy has been provided \cite{PW92}.

The Migdal theorem \cite{Mi58} contains a similar statement that for an electron-phonon coupling, higher-order vertex corrections vanish in orders of the ratio of the phonon frequency to the Fermi energy. Violations of this theorem appear if the magnon frequency becomes large \cite{HLB76,IOS92} or non-adiabaticity leading eventually to a polaron collapse \cite{AGM87}. For heavy fermion systems the Migdal theorem is not valid \cite{W99} and near the magnetic boundary the quasiparticle spectra is different from the Eliashberg theory \cite{Mon03} which means the applicability of RPA calculations in high-$T_c$ superconductivity \cite{GPS95,GPS95a} is questionable.

It is therefore desirable to have a simple criterion when vertex corrections vanish in the high-density limit. Here we provide an argument from simple scaling for interacting Fermi systems which shows that the power exponent of the interaction and the dimensionality of the system together with the exponent of the relation between Fermi energy and momentum, determines the expansion of the vertex corrections in terms of the Fermi momentum. First we drive an exact scaling law combining the dimensionality, the form of Fermi energy, and the momentum behavior of the potential into a condition when RPA is exact. As an application, we calculate the structure factor and pair correlation function for a wire of Fermions and compare the result with recent Quantum Monte Carlo simulation data \cite{VP18,VBMP17} to check the proposed high-density expansion. 

An exact scheme for many-body correlations based on the variational technique by Hedin \cite{H65} is used in various applications. This scheme allows a systematic numbering of Feynman diagrams \cite{Mol05} and has been solved exactly in zero dimensions \cite{PH07}. The scheme provides systematic vertex corrections beyond the GW approximation \cite{ScG98} giving evidence of a convergence of the expansion. It is as well useful to describe spin-dependent interactions \cite{AB09}. For an overview about recent numerical methods to solve GW approximations and corresponding Bethe-Salpeter equations see
\cite{LJWM16}.

We will use this Hedin scheme to analyze the high-density limit. To recall the basic formulas,
let us consider the causal propagator as the many-body averaging of the time-ordered product ${\cal T}$ of creation and annihilation operators
$G=\frac 1 i\langle {\cal T} \Psi(1)\Psi^+(2)\rangle$ where arguments denote cumulatively space,time,... coordinates.

\begin{figure}
\centerline{\includegraphics[width=6cm]{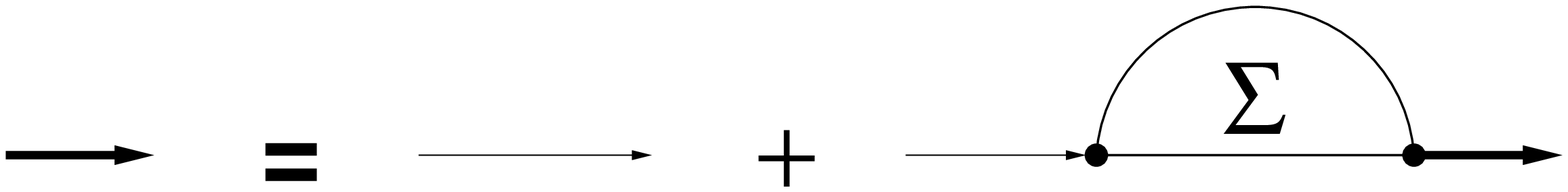}}
\centerline{\includegraphics[width=6cm]{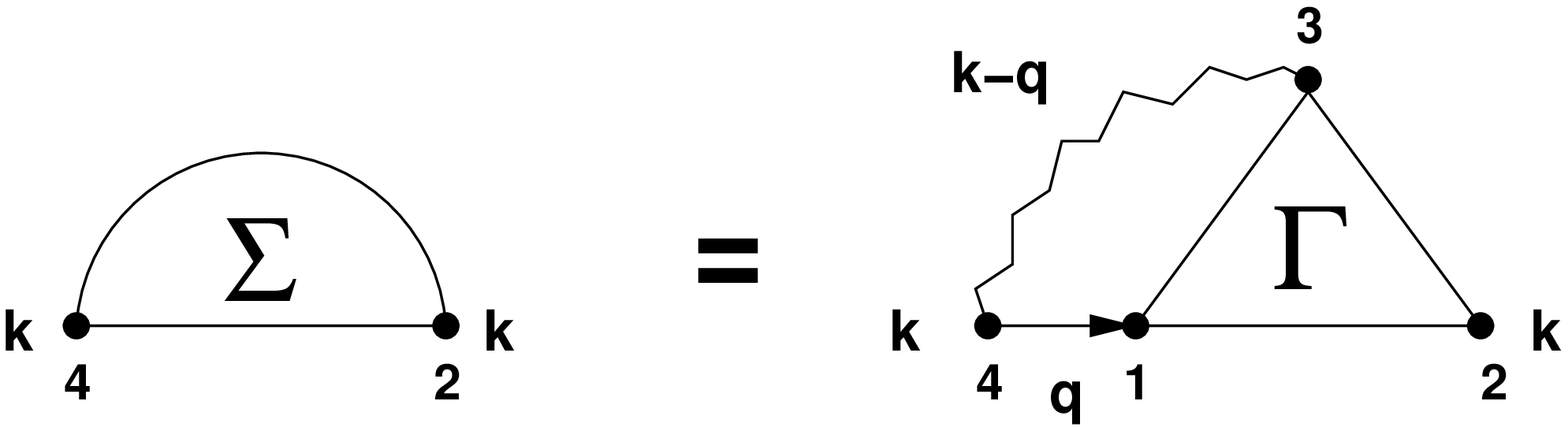}}
\centerline{\includegraphics[width=4cm]{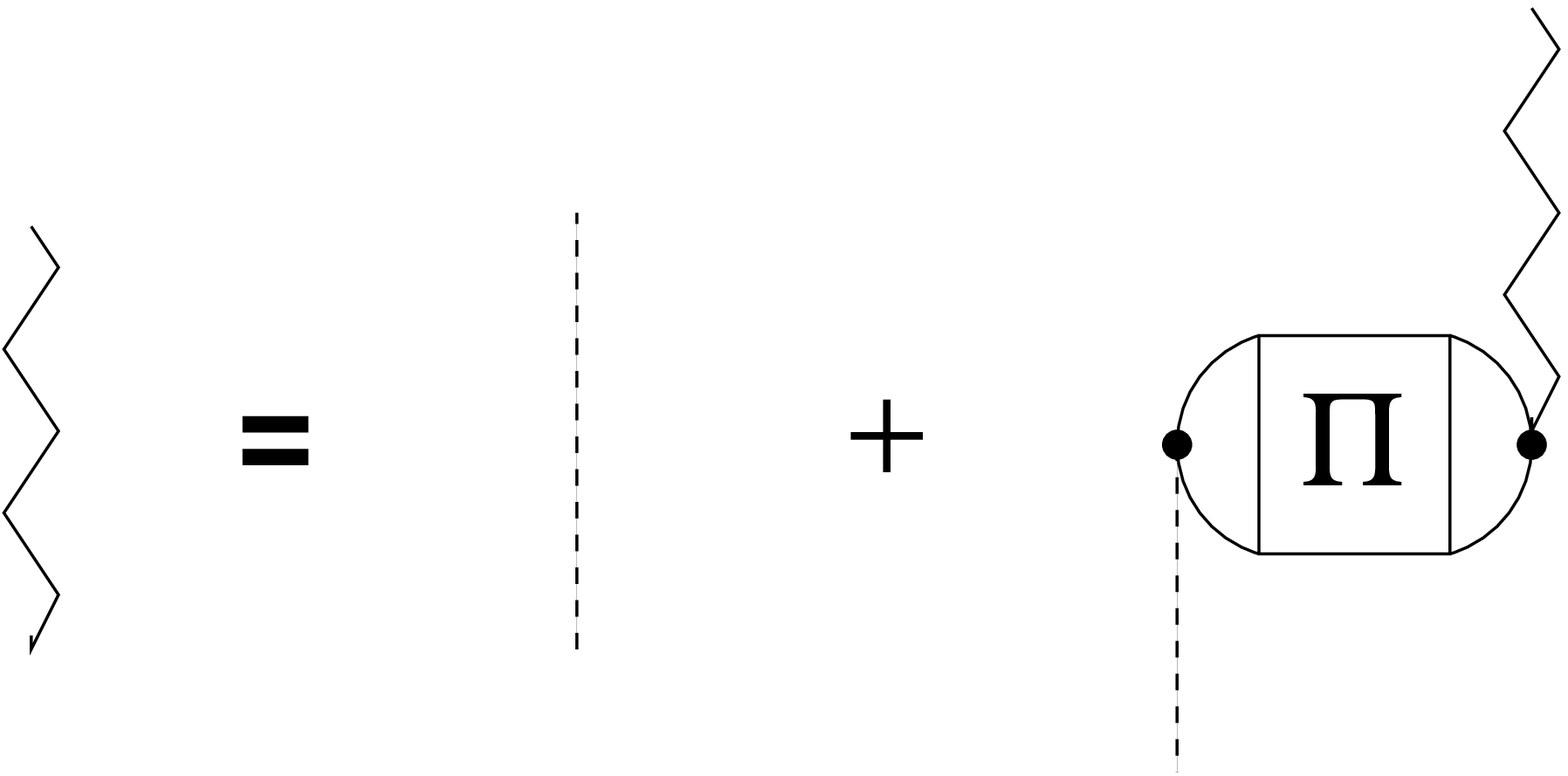}}
\centerline{\includegraphics[width=6cm]{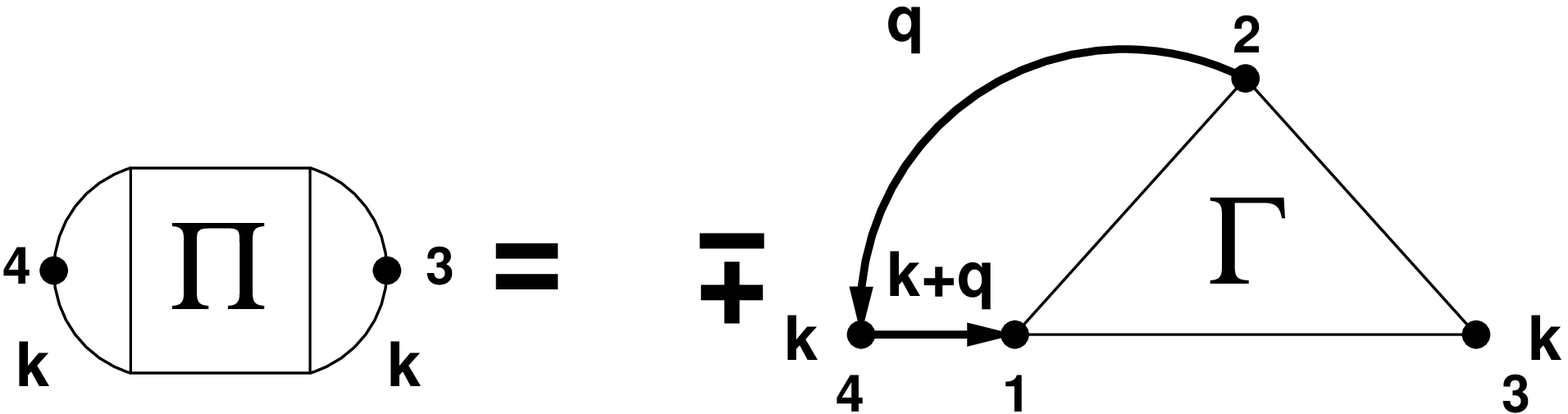}}
\centerline{\includegraphics[width=6cm]{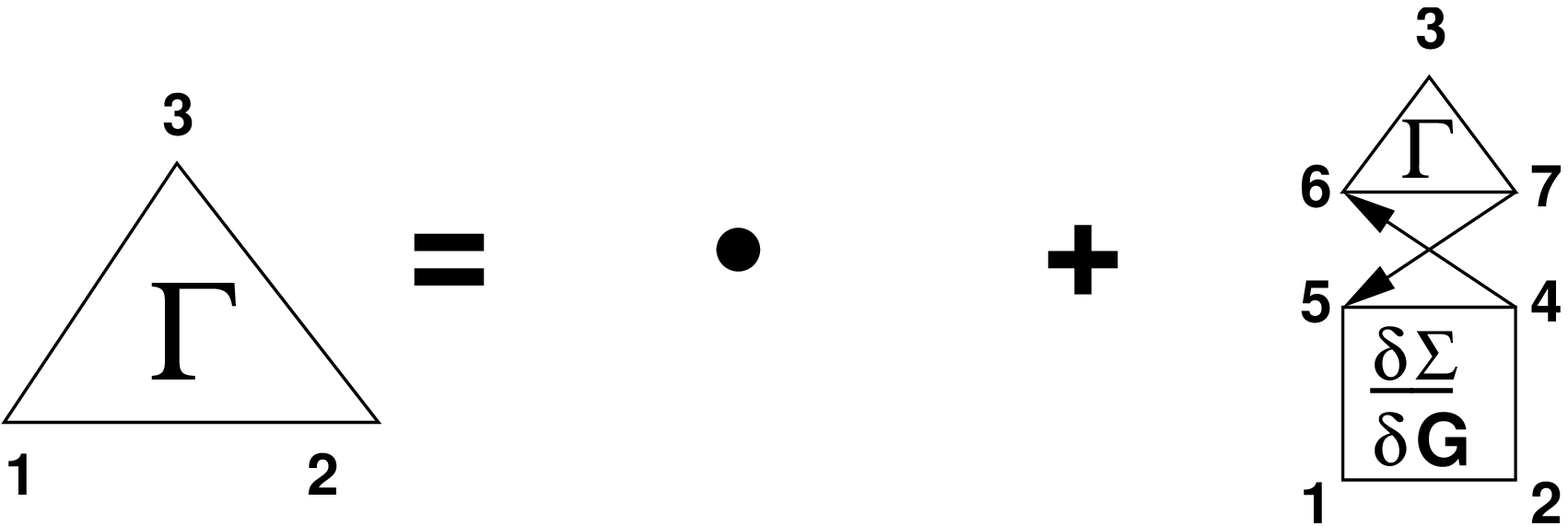}}
\caption{\label{diag}  The 5 Hedin equations where the thin line is $G_0$. The numbers give the space-time variable in general and the letters denotes the frequency-momentum ones after Fourier transform in equilibrium. Please note that the $\mp$ sign for Fermi/Bosons will be absorbed into closed propagator lines when expanded to get the standard Feynman diagrams.}
\end{figure}

The Hedin equations consist first of the Dyson equation for this
causal propagator
\be
G(1,2)=G_0(1,2)+G_0(1,3)\Sigma(3,4)G(4,2)
\ee
where about double occurring indices will be integrated. The selfenergy $\Sigma$ is given in terms of the three-point vertex function $\Gamma$ and the screened potential $W$ as
\be
\Sigma(1,2)=G(1,3)W(1,4)\Gamma(3,2,4).
\label{s}
\ee
The screened potential $W$ in turn is determined by the interaction potential $V$ and the polarization $\Pi$ in a RPA-like equation
\be
W(1,2)=V(1,2)+V(1,3)\Pi(3,4)W(4,2).
\label{W}
\ee
The polarization can be expressed by the vertex function similar to (\ref{s})
as
\be
\Pi(1,2)=\mp G(1,3)G(4,1)\Gamma(3,4,2)
\label{P}
\ee
for fermions/bosons respectively. Compared to (\ref{s}) the last two variables are interchanged. To close the above equation system one can express now the vertex function as the variation of the selfenergy with respect to the propagator
\be
\Gamma(1,2,3)
%&=&\delta_{12}\delta_{13}+{\delta {\bar \Sigma}_{12}\over \delta {\bar U}_{33}}
%\nonumber\\
=\delta_{12}\delta_{13}+{\delta {\Sigma}(1,2)\over \delta G(4,5)} G(4,6) \Gamma(6,7,3) G(7,5).
\label{g123}
\ee
This set of five equations are exact in equilibrium and non-equilibrium and can be shown by variational technique \cite{Kl17}. In equilibrium the functions above become dependent on the difference of the space and time coordinates and a Fourier transformation leads to actually a factorization in (Matsubara) frequencies and momentum with one integration remaining for closed loops.
In short we write $p=(\omega,\V p)$. Pictorially we illustrate the set of equations in figure~\ref{diag}.

Now we analyze the exact Hedin equations with respect to their density dependence. Therefore we scale explicitly all momentum and energy variables in terms of Fermi momentum. This scaling is not providing a dimensionless quantity. We only extract the dependence on the Fermi momentum. The Fermi momentum as the radius of Fermi sphere is related to the density via $n\sim p_{\rm f}^d$. Very often the expansions are given in terms of the Wigner-Seitz radius which describes the radius of a sphere containing one particle. Therefore the scaling with the density is $r_s\sim n^{-1/d}$ and one has $p_{\rm f}\sim 1/r_s$. The high-density limit is the high-Fermi-momentum limit or the small-$r_s$ limit.

An integration over an internal momentum and energy is proportional to $p_{\rm f}^{d+\beta}$ where $d$ is the dimension of the system and we assume a general relation between Fermi energy and momentum of $\epsilon_{\rm f}\sim p_{\rm f}^\beta$. For quadratic dispersion we have of course $\beta=2$ but one might think on non-Fermi liquids as well. The used potential is supposed to have an $~q^{-\alpha}$ dependence which leads to the scaling
$
V=\{\ov V\} \, p_{\rm f}^{-\alpha}
$
where we will denote the scaled function with the bracket $\{\}$ and an over-line.
The propagator scales as inverse energy $G=\{\ov G\}\, p_{\rm f}^{-\beta}$ and we have in (\ref{W}) only one internal momentum-energy integration, see figure~\ref{diag}, such that we obtain for (\ref{W})
\be
W&=&\{\ov V\} \, p_{\rm f}^{-\alpha}+\{\ov V\} \,  \Pi \, W \, p_{\rm f}^{d+\beta-\alpha}.
\ee
This implies the scaling $W=\{\ov W\} p_{\rm f}^{-\alpha}$ and $\Pi=\{\ov \Pi\}\, p_{\rm f}^{\alpha-\beta-d}$. Equation (\ref{P}) also contains  only one internal integration such that one gets
\be
\Pi&=&\{\ov G\}\, \{\ov G\}\,  \Gamma \, p_{\rm f}^{d-\beta}
\ee
which let us conclude $\Gamma=\{\ov \Gamma\}\, p_{\rm f}^{\alpha-2 d}$.
Analogously one sees for (\ref{s}) and (\ref{g123})
\be
\Sigma&=&\{\ov G\} \, \{\ov W\}\,  \Gamma \, p_{\rm f}^{d-\alpha}\nonumber\\
G&=&\{\ov G_0\}\, p_{\rm f}^{-\beta}+\{\ov G_0\}\, \Sigma \,  \{\ov G\} \, p_{\rm f}^{d-\beta}.
\ee
Therefore this implies $\Sigma=\{\ov \Sigma\}\, p_{\rm f}^{-d}$ which fits the fourth equation.
The equation for the vertex (\ref{g123}) finally scales as
\ba
\Gamma&=1\!+\!{\delta \{\ov \Sigma\}\{\ov W\} \Gamma \over \delta \{\ov G\}}  \{\ov G\}\Gamma \{\ov G\}   p_{\rm f}^{d\!-\!\beta\!-\!\alpha}
=1\!+\!\{...\}\Gamma^2 p_{\rm f}^{d-\beta-\alpha}\nonumber\\&
=1+O(%\left (
p_{\rm f}^{d-\beta-\alpha})%\right )
.
\label{gexp}
\end{align}
Here we have used in (\ref{g123}) the expression for the self energy (\ref{s}) before scaling. This equation (\ref{gexp}) for the vertex shows how further iterations scale with the orders of the Fermi momentum.
In other words, the vertex function becomes unity in first order of the power of Fermi momentum $p_{\rm f}^{d-\beta-\alpha}$. In case that $d-\beta-\alpha<0$  the RPA is consequently exact in the high-density limit.

As an illustrative example let us calculate the static structure factor
for Fermions in one-dimension
\be
S(q)&=&-{1\over n\pi}\int\limits_0^\infty d \omega {\rm Im} {\Pi \over 1-V_q \Pi}
\nonumber\\
&=&-{1\over n\pi}\int\limits_0^\infty d \omega  {{\rm Im}\Pi \over (1-V_q {\rm Re}\Pi)^2+(V_q{\rm Im} \Pi)^2}.
\label{Sq}
\ee
The pair-correlation function is obtained from the structure factor as
\be
g(r)=1+{1\over n}\int\limits_{-\infty}^\infty {dq \over 2\pi \hbar }{\rm e}^{{i\over \hbar} q r} (S(q)-1).
\label{gr}
\ee

\begin{figure}
\centerline{\includegraphics[width=8cm]{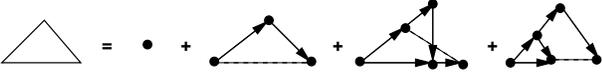}}
\caption{\label{vert}  The expansion of the vertex function (\ref{g123}) up to first-order interaction.}
\end{figure}

We first consider the expansion with respect to the interaction potential and will see that it agrees with the high-density limit. To calculate the vertex equation (\ref{g123}) we need the variation of the selfenergy (\ref{s}). The latter one becomes together with the screened potential (\ref{W})
\be
{\delta \Sigma(1,2)\over \delta G(4,5)}=\delta_{14}\delta_{35}W(1,6)\Gamma(3,2,6)+O(V^2)
\ee
such that the vertex (\ref{g123}) gets the expansion as illustrated in figure~\ref{vert}.

\begin{figure}
\centerline{\includegraphics[width=8cm]{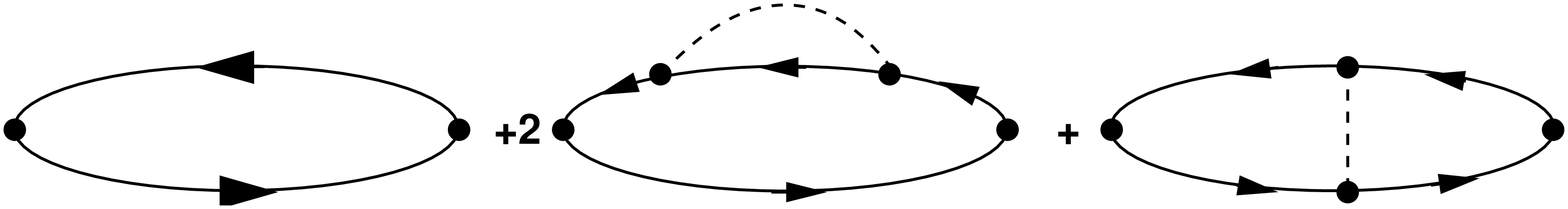}}
\caption{\label{pol}  First-order expansion of the polarization function (\ref{P}) in terms of interaction when using the vertex of figure~\ref{vert}.}
\end{figure}

The resulting polarization (\ref{P}) is plotted in figure~\ref{pol} where the first term is the RPA result
\be
\Pi_0(q,\omega)=\int{dp\over 2 \pi \hbar} {f_{p+\frac q 2}-f_{p-\frac q 2}\over \epsilon_{p+\frac q 2}-\epsilon_{p-\frac q 2}-\hbar \omega-i0}
\ee
which becomes for zero temperature
\be
{\rm Im} \Pi_0&=&-{m g_s\over 2 \hbar q}\Theta(\omega-|\omega_-|)\Theta(|\omega_+|-\omega),\nonumber\\
{\rm Re} \Pi_0&=&{m g_s\over 2 \pi \hbar q} \ln{\left |{\omega^2-\omega_-^2\over \omega^2-\omega_+^2}\right |}
\label{p0t}
\ee
with $\hbar \omega_\pm={q\over m}(\frac q 2 \pm p_{\rm f})$, the Fermi momentum $p_{\rm f}=\hbar k_{\rm f}$, and spin degeneracy $g_s$.

Expanding the response function in terms of the coupling parameter $r_s$ requires to consider the self-energy and vertex corrections beyond $\Pi_0$ which we denote as $\Pi_0+\tilde \Pi$ plotted in figure~\ref{pol}.
If we now expand the integrand of (\ref{Sq}) with respect to the interaction which is the expansion in terms of the coupling parameter $r_s$, we get
\ba
{\rm Im}{\Pi\over 1\!-\!V\Pi}={\rm Im} \Pi_0\!+\!
{\rm Im} \tilde \Pi\!+\!2 V {\rm Im}\Pi_0{\rm Re}\Pi_0\!+\!O(V^2).
\label{exp}
\end{align}
We see that the correction to the polarization contributes to the same order in the potential as the expansion of the denominator of (\ref{Sq}). We will denote with subindex $_1$ this first part of expansion in interaction to distinguish from the vertex correction signed by the subscript $_v$.

Let us inspect the corrections $\tilde \Pi=\Pi_{\rm se}+\Pi_v$ more in detail.
The selfenergy correction $\Pi_{\rm se}$ as the second part of figure \ref{pol} is given by
\be
\Pi_{\rm se}=\sum_{k,p}\frac{V_q(k-q)(f_{k}-f_{k+q})(f_{p}-f_{p+q})}{(\hbar \omega+(\epsilon_k-\epsilon_{k+q}))^2}
\ee
and can be written as the frequency derivative $\Pi_{\rm se}\sim \partial_\omega (\omega-pq/m)^{-1}$ such that it does not contribute to (\ref{Sq}). In fact, due to (\ref{exp}) the contribution of $\Pi_{\rm se}$ goes to zero by integrating over the imaginary axis as has been done in \cite{RMSP12}. The only contributing part is the last one of figure~\ref{pol} which is the vertex correction $\Pi_{v}$ given by
\be
\Pi_{v}(q,\omega)&=&-\int {d p d k\over (2\pi \hbar)^2}
V_{p-k}{f_{p+\frac q 2}-f_{p-\frac q 2}\over \epsilon_{p+\frac q 2}-\epsilon_{p-\frac q 2}-\hbar \omega-i0}
\nonumber\\
&&\times
{f_{k+\frac q 2}-f_{k-\frac q 2}\over \epsilon_{k+\frac q 2}-\epsilon_{k-\frac q 2}-\hbar \omega-i0}.
\label{pvert}
\ee

For contact potentials we have $\Pi_{v}=-V\Pi_0^2$ and ${\rm Im}\Pi_{v}=-2 V {\rm Im}\Pi_0{\rm Re}\Pi_0$ and consequently an exact cancellation of the first-order term in (\ref{exp}) and we get the non-interacting structure factor.

For finite-range potentials we also expect a compensation with some small effect remaining which we are going to calculate now for the example of a cylindrical wire with a smoothed real-space interaction potential $v(r) \varpropto(r^2+b^2)^{-1/2}$ and its Fourier transform
$V(q)= 2~\frac{e^2}{4 \pi \epsilon_{0}}~K_{0}(b q)$, where $K_{0}$ is the modified Bessel function of
second kind and $b$ being the width of the wire. 
The imaginary part of (\ref{pvert}) is easily found by partial decomposition
and integrating over $p$.
%\ba
%&{\rm Im}\Pi_{v}={m^2\over 4\pi \hbar^2 q^2}
%\left [f_{{m\omega\over q}-{q\over 2}}-f_{{m\omega\over q}+{q\over 2}} \right ]
%\int\limits_{-\infty}^\infty {d k\over k} V_k
%\nonumber\\
%&\left [f_{{m\omega\over q}-{q\over 2}-k}-f_{{m\omega\over q}+{q\over 2}-k}
%-f_{{m\omega\over q}-{q\over 2}+k}+f_{{m\omega\over q}+{q\over 2}+k} \right ].
%\end{align}
The imaginary part of $\Pi_0$ of (\ref{p0t}) factors out and for zero-temperature and using $V_k=V_{-k}$ we obtain
\be
{\rm Im}\Pi_{v}=\!-{\rm Im}\Pi_0 {m g_s\over \pi \hbar q}
\!\left [
\int\limits_{{m\hbar \over q} (\omega-|\omega_+|)}^{{m\hbar \over q} (\omega-|\omega_-|)}
\!\!+\!\!
\int\limits_{{m\hbar \over q} (\omega+|\omega_+|)}^{{m\hbar \over q} (\omega+|\omega_-|)}
\right ] \! {d k\over k} V_k
\label{Pv}
\ee
with $\hbar \omega_\pm ={qp_{\rm f}\over m}\pm {q^2\over 2m}$. For the explicit calculation it is convenient to use $x=q/2p_{\rm f}$ and $z=m \omega/2p_{\rm f}^2x$ to get for the structure factor (\ref{Sq}) with (\ref{exp})
\be
S(q)=S_0(q)+S_1(q)+S_{v}(q)
\label{S1v}
\ee
with the free one
\be
S_0(q)=\frac 1 2 \int\limits_{|1-x|}^{|1+x|} d z
=x \Theta(1 - x)+\Theta(x-1 ).
\label{S0}
\ee
The next term of (\ref{exp}) is the one from the expansion of the denominator, $2 V {\rm Im} \Pi_0 {\rm Re} \Pi_0$ which reads \cite{VBMP17} for $x<1$
\be
S_1(q)=-v(q)\frac{g_s^2\;r_s}{\pi^2\; x}
\bigg[
\left |1 \!-\! x \right |\ln\left |1\!-\!x\right |
%\nonumber\\
+\left (1 \!+\! x\right ) \ln{\left (1 \!+\! x\right )}
\bigg],\nonumber\\
\label{S1}
\ee
and an additional term  $-2 x \ln x$  for $x>1$. In the small limit of $x$, the $S_1(q)$ takes the simpler form
\be
S_1(q)=-v(x\to0)\frac{g_s^2\;r_s}{\pi^2}x. 
\ee
%\ba
%S_1(q)
%%=&
%%\frac {g_s^2 r_s} {2\pi^2} {{v}(x)\over x}\int\limits_{|1-x|}^{|1+x|} d z
%%\ln{{z^2-(1-x)^2\over (1+x)^2-z^2}}
%%\nonumber\\
%=&
%-v(q)\frac{g_s^2\;r_s}{\pi^2\; x}
%\bigg[
%\left |1 \!-\! x \right |\ln\left |1\!-\!x\right |
%%\nonumber\\
%+\left (1 \!+\! x\right ) \ln{\left (1 \!+\! x\right )}
%\bigg]
%\label{S1}
% \end{align}.
% \nonumber\\
%=&    -v(q\to0)\frac{g_s^2\;r_s}{\pi^2}x ,~ x\ll 1.
%\ba
%S_{1}(q)=-v(q)\frac{g_s^2\;r_s}{\pi^2\; x}
%\bigg[
%&\left (x \!-\!1 \right )\ln\left (x\!-\!1\right )
%+\left (1 \!+\! x\right ) \ln{\left (1 \!+\! x\right )}
%\nonumber\\
%&-2 x \ln x\bigg].
%\label{S1a}
%\end{align}
Here and in the following we use $V(q)=v(q) e^2/4 \pi \epsilon_0$. 

The vertex correction (\ref{Pv}) with (\ref{exp}) leads to the part
\be
S_{v}(q)=-r_s {g_s^2\over 2 \pi^2 x} \int\limits_{|1-x|}^{|1+x|} d z\,
(
\int\limits_{z-|1+x|\over 2}^{z-|1-x|\over 2}+
\int\limits_{z+|1+x|\over 2}^{z+|1-x|\over 2}
)
{d \bar x\over \bar x} v(\bar x)
\ee
where we will now interchange the order of integrations. The first integral vanishes and the second one leads to
\ba
S_{v}(q)=-r_s {g_s^2\over \pi^2 x}
&\left [
\left (
(1+x)\int\limits_{1}^{1+x}-
(1-x)\int\limits_{1-x}^{1}
\right )
{d \bar x\over \bar x} {v}(\bar x)
\right .
\nonumber\\
&\left .
+
\left (
\int\limits_{1-x}^{1}-
\int\limits_{1}^{1+x}
\right )
d {\bar x} v(\bar x)
\right ]
\label{Sv}
\end{align}
for $x<1$ and similarly for $x>1$
\ba
S_{v}(q)=-r_s {g_s^2\over \pi^2 x}
&\left [
\left (
(1+x)\int\limits_{x}^{1+x}-
(x-1)\int\limits_{x-1}^{x}
\right )
{d \bar x\over \bar x} {v}(\bar x)
\right .
\nonumber\\
&\left .
+
\left (
\int\limits_{x-1}^{x}-
\int\limits_{x}^{1+x}
\right )
d {\bar x} v(\bar x)
\right ].
\label{Sva}
\end{align}
The explicit integrals appearing in (\ref{Sv}) and (\ref{Sva}) can be solved analytically and are given explicitly in the Appendix.
Especially interesting is the small-$q$ expansion of (\ref{Sv}) like (\ref{S1}) which yields for any potential
\be
S_{v}(q)=-{2g_s^2 r_s \over \pi^2 } v(x\to 0) x+O\left (x\right )^2.
\ee

\begin{figure}
\centerline{\includegraphics[width=8cm]{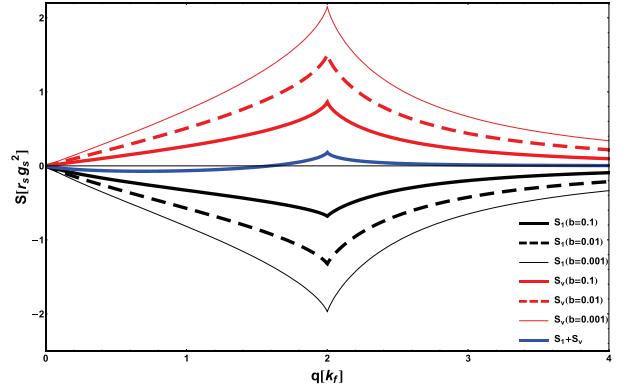}}
\caption{\label{vertex}  (\footnotesize{Color online}) The first-order corrections in $r_s$ to the structure factor according to the denominator (\ref{S1}), (\ref{S1a}) (black, lower lines) and the vertex corrections (\ref{Sv}), (\ref{Sva}) (red, upper lines) for a cylindrical potential. The sum of both corrections is nearly independent of $b$ (blue, middle line) and visually not distinguishable.}
\end{figure}

In figure~\ref{vertex} we compare the first-order corrections (\ref{S1}) with the vertex correction (\ref{Sv}). As expected there is a large compensation of both corrections. Interestingly the remaining sum of both corrections $S_1$ and $S_v$ is independent of the width parameter as can be seen analytically  and we get for $x<1$
\ba
S_1+S_{v}=&
\frac{r_s g_s^2}{\pi ^2
   x}
\left \{
(1+x) \ln(1+x)[2+ \ln {{x^2\over 1+x}}]
\right .
\nonumber\\
&\left .
+
|1-x| \ln|1-x|[2+ \ln {{x^2\over |1-x|}}]
\right\},
\label{S1v}
\end{align}
and for $x>1$ an additional term $2 x \ln x [2+\ln x]$ appears.

 We conclude that for contact interactions up to second order in the interaction or $r_s$ parameter, the structure factor (\ref{Sq}) is the interaction free-one (\ref{S0}) valid approximately also for finite short-range interactions.
This provides the non-interacting pair-correlation function (\ref{gr})
\be
g_0(r)=1+ 2 {\cos{\left ({2 k_{\rm f} r}\right )}-1\over \left ({2 k_{\rm f} r}\right )^2}.
\label{g0}
\ee
The first-order high-density ($r_s$) corrections due to (\ref{S1v}) are
\be
g_{1}(r)+g_{v}(r)=-{2 r_s g_s^2\over 9 \pi^2}(3+\pi^2) \left ({r k_{\rm f}}\right )^2+o(r k_{\rm f})^4
\label{30}
\ee
and we see that the artifact of RPA to provide negative pair-correlation functions at small distances is cured due to the cancellation of vertex and RPA corrections.

Let us return to the high-density limit. In one dimensional systems, $d=1$, for quadratic dispersion, $\beta=2$, and contact interaction, $\alpha=0$, according to (\ref{gexp}) we have
%\be
$\Gamma=1+O(p_{\rm f}^{-1})$
%\label{gf}
%\ee
and the RPA is the exact limit as we have shown above. For the cylindrical potential we have the expansion \be
 v(q)=\left\{%
\begin{array}{lcl}
\label{ClyPoten}
    -\gamma+\ln(2)-\ln (b q) &\text{for} & \hbox{ bq $\rightarrow$ 0} \\
   e^{-bq}\sqrt{\frac{\pi}{2bq}}&\text{for} & \hbox{ bq $\rightarrow$ $\infty$.} \\
\end{array}%
\right.\ee
therefore $K_0[bq]$ has no scale of Fermi momentum i.e $\alpha=0$ which is also clear from the Fourier transform of $1/x$. Therefore we have $\Gamma=1+o(p_{\rm f}^{-1})$ for $\beta=2$ and $d=1$ rendering the RPA exact in the high-density limit. In detail we have seen that up to the same order the structure factor becomes the non-interacting one. 

\begin{figure}
\centerline{\includegraphics[width=8cm]{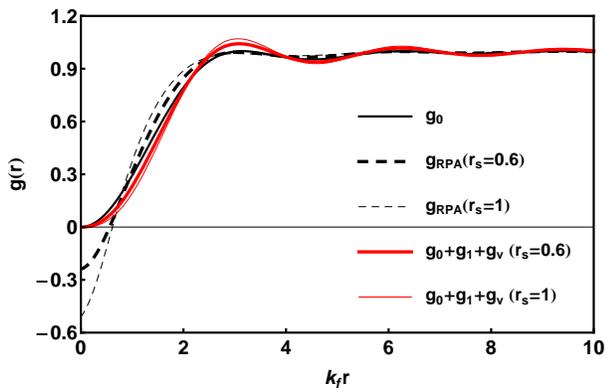}}
\caption{\label{grp} (Color online) The interaction-free pair-correlation function, $g_0$ which is the high-density limit (\ref{g0}) together with the ones from RPA (\ref{Sq}), and the first-order expansion (\ref{S1v}) in cylindrical potential with thickness parameter $b=0.1$.}
\end{figure}

\begin{figure}[!t]
\centerline{\includegraphics[width=8.cm]{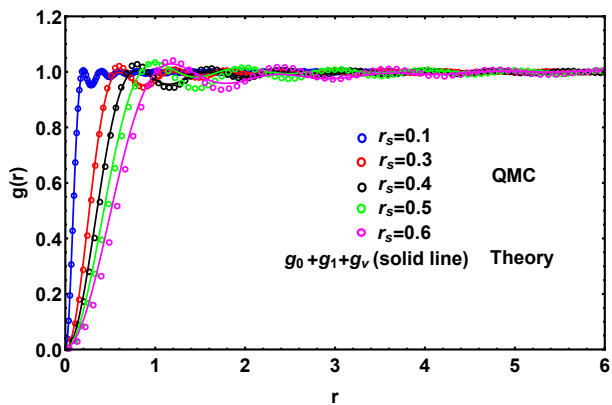}}
\caption{\label{grp_qmc}(Color online) The pair-correlation function $g(r)$ with first order in $r_s$ including the vertex correction is plotted as a function of $r$ at several densities, and are compared with recent quantum Monte Carlo (QMC) simulation data \cite{VP18}.}
\end{figure}

We use (\ref{exp}) in (\ref{Sq}) and get the structure factor $S(q)$ for cylindrical wire and employing (\ref{gr}) to obtain $g(r)$ with no negative values of the pair correlation function at small distances (\ref{30}) as seen in the figure~\ref{grp}.
In this figure the pair correlation function is plotted in various approximations. We see that the RPA leads to the known negative values at small distances since the exchange (vertex) correction are not included. The interaction potential alone in the denominator of RPA is not justified neither by the small interaction nor the high-density limit.
This artifact is removed if we include the first-order exchange correction to the RPA in the high density limit. In figure~\ref{grp_qmc} we compare our pair correlation function with the Quantum Monte Carlo (QMC) simulation  \cite{VP18}. The details of QMC calculations are mentioned in the reference \cite{Needs10,Lee11}. One sees a good agreement for small distances as well as for the oscillations at larger distances. The high-density limit of RPA can only be approached numerically but not analytically.  

To summarize we have shown by scaling of the exact Hedin equations with respect to Fermi momentum that vertex corrections behaves as $p_{\rm f}^{d-\beta-\alpha}$ with dimensionality $d$, particle dispersion $\epsilon_p\sim p^\beta$, and potential $V_q\sim q^{-\alpha}$. That means for $\beta+\alpha>d$ the vertex corrections vanish in the high-density limit. This scaling relation also answers the question under which conditions one might expect an exception of the Migdal rule mentioned in the introduction. For 3D Systems ($d=3$) and Coulomb interaction ($\alpha=2$) we see that the dispersion relation between momentum and energy ($\epsilon_f\sim p_{\rm f}^\beta$) has to have a value $\beta>1$ in order to allow the vanishing vertex correction in the high-density limit. For non-Fermi liquids with $\beta=1$ we see that the vertex corrections do not vanish and we have different properties like anomalous transport and violation of Migdal's rule. 

We have discussed the structure factor and pair correlation function and show that there is a cancellation of vertex corrections and the RPA denominator maintaining the free result up to second-order interaction for contact interactions. This result is in line with the general observation that some many-body effects like phase transitions get reduced when using approximations beyond RPA. The high-density limit is shown to be the same as the interaction-free result consistent with the expansion in the vertex.
For a finite-range potential, the compensation of vertex and RPA corrections renders the structure factor  independent of the width parameter and repairs the pair-correlation function to be positive at small distances.

The authors (VA and KNP) acknowledge the financial support by National Academy of Sciences of India. KM like to thank for the support for his visit to Panjab university by DFG and Indian National Science Academy.

\appendix*
\section{Expansion in terms of the width of cylindrical wire}

To get the small $b$ expansion of  $S_1(q)$ we simply expand $v_q$ for a cylindrical wire. The first order correction to the structure factor for $x<1$ is given by 
\ba
S_1(q)=&
2 \left(\ln \left(\frac{b
   q}{2}\right)+\gamma \right) \frac{g_s^2\;r_s}{\pi^2\; x}
\bigg[
\left (1 - x\right )\ln\left (1-x\right )\nonumber\\
+&\left (1 + x\right ) \ln{\left (1 + x\right )}
\bigg]
\label{AS1}
 \end{align}
Similarly, for $x>1$ one obtains
\be
S_{1}(q)=&2 \left(\ln \left(\frac{b
   q}{2}\right)+\gamma \right)\frac{g_s^2\;r_s}{\pi^2\; x}
\bigg[
\left (x -1 \right )\ln\left (x-1\right )
\nonumber\\
&+\left (1 + x\right ) \ln{\left (1 + x\right )}
-2x \ln x\bigg].
\label{S1a}
\ee

%\begin{figure}[h]
%\centerline{\includegraphics[width=8cm]{gr_final.eps}}
%\caption{\label{grp1} Pair correlation function in RPA, RPA with vertex and self energy correction and its systematic expansion in power of $r_s$. Corrections to order $r_s$ to $g(r)$ are denoted by $g_1$ and $g_v$.}
%\end{figure}

The finite-$b$ results
for the cylindrical potential are
\ba
\int\!\! 2 K_0(b t) {d t\over t}&=-\frac{1}{2} G_{1,3}^{3,0}\left(\frac{b t}{2},\frac{1}{2}|
\begin{array}{c}
 1 \\
 0,0,0 \\
\end{array}
\right)
\nonumber\\
\int\!\! 2 K_0(b t) {d t}&=
\pi  t [\pmb{L}_{-1}(b t) K_0(b t)+\pmb{L}_0(b t) K_1(b t)]
\label{intSv}
\end{align}
in terms of the modified Struve function $L_n(x)$ and the Meijer $G$ function. It is noted that it is easier to take the limit $b\rightarrow 0$ from  (\ref{Sv}) and (\ref{Sva}) by expanding $K_0(b t)$ appearing in the integrand since after integration the limit of Meijer function for $b\rightarrow 0$ does not exist. 
The analytical expression for finite $b$ of the vertex correction for $x<1$ is given by
\ba
&S_v(q)=-\frac{r_s g_s^2}{\pi^2 x}
   \bigg\{ {x-1\over 2}
   G_{1,3}^{3,0}\left(\frac{b-bx}{2},\frac{1}{2}|
\begin{array}{c}
 1 \\
 0,0,0 \\
\end{array}
\right)
\nonumber\\& 
\!+\!
   G_{1,3}^{3,0}\left(\!\frac{b}{2},\frac{
   1}{2}|
\begin{array}{c}
 1 \\
 0,0,0 \\
\end{array}
\!\right)
\!-\!{x\!+\!1\over 2}
   G_{1,3}^{3,0}\left(\!
\frac{bx+x}{2},\frac{1}{2}|
\begin{array}{c}
 1 \\
 0,0,0 \\
\end{array}
\!\right)
%\nonumber\\& 
\nonumber\\& \!-\!\pi
(x\!+\!1)
   \left [
\pmb{L}_{-1}(b x\!+\!b) K_0(b
   x\!+\!b)
\!+\!\pmb{L}_0(b x\!+\!b) K_1(b
   x\!+\!b)
\right ]
\nonumber\\&
\!+\!\pi (x\!-\!1) \left [
\pmb{L}_{-1}(b\!-\!b x)
   K_0(b\!-\!b x)\!+\!\pmb{L}_0(b\!-\!b x) K_1(b\!-\!b
   x)
\right ]\nonumber\\
   & +2\pi \left [ \pmb{L}_{-1}(b) K_0(b)+
   \pmb{L}_0(b) K_1(b)\right ]
\bigg\},
\end{align}
and for $x>1$ as,
\ba
&S_v(q)=-\frac{r_s g_s^2}{\pi ^2 x}
   \bigg\{
\frac{1-x}{2} 
   G_{1,3}^{3,0}\left(\frac{b x-b}{2},\frac{1}{2}|
\begin{array}{c}
 1 \\
 0,0,0 \\
\end{array}
\right)
\nonumber\\
& \!+\!x G_{1,3}^{3,0}\left(\frac{b
   x}{2},\frac{1}{2}|
\begin{array}{c}
 1 \\
 0,0,0 \\
\end{array}
\right)\!-\!{x\!+\!1\over 2}
   G_{1,3}^{3,0}\left(\frac{b x\!+\!b}{2},\frac{1}{2}|
\begin{array}{c}
 1 \\
 0,0,0 \\
\end{array}
\right)
\nonumber\\&
\!-\!\pi (x\!+\!1)
\left [
\pmb{L}_{-1}(b x\!+\!b)
   K_0(b x\!+\! b)\!+\!\pmb{L}_0(b x\!+\! b)K_1(bx\!+\!b)
\right ]
\nonumber\\&
\!-\!\pi (x\!-\!1) 
\left [
\pmb{L}_{-1}(b\!-\!b x) K_0(b x\!-\!b)
\!-\!\pmb{L}_0(b\!-\!b x) K_1(b x\!-\!b)
\right ]
\nonumber\\&
+2\pi x \left [
\pmb{L}_{-1}(b x)K_0(b x)+\pmb{L}_0(b x) K_1(b
   x)\right ]
\bigg \}.
\end{align}

It is also noted that the negativeness of the pair correlation function also depends on the thickness of the wire. For infinitesimally small thickness, the pair correlation function remains positive.
%as seen in figure~\ref{grp1}.

%\bibliography{bose,kmsr,kmsr1,kmsr2,kmsr3,kmsr4,kmsr5,kmsr6,kmsr7,delay2,spin,spin1,refer,delay3,gdr,chaos,sem3,sem1,sem2,short,cauchy,genn,paradox,deform,shuttling,blase,spinhall,spincurrent,tdgl,pattern,zitter,isospin,quench,hubbard,march}
%\bibliographystyle{prsty}

\end{document}